\documentclass{article}

\usepackage{spconf}
\usepackage{amsmath, graphicx, booktabs}
\usepackage{wrapfig, url}
\usepackage{threeparttable}

\title{AISHELL-3: A Multi-speaker Mandarin TTS Corpus and the Baselines}
\name{Yao Shi$^{1,2}$, Hui Bu$^{3}$, Xin Xu$^{3}$, Shaoji Zhang$^{3}$, Ming Li$^{1,2}$ }

\address{
    $^{1}$ School of Computer Science, Wuhan University, Wuhan, China  \\
    $^{2}$ Data Science Research Center, Duke Kunshan University, Kunshan, China \\
    $^{3}$ Beijing Shell Shell Technology Co., Ltd \\
    \url{ming.li369@dukekunshan.edu.cn}\\
}

\begin{document}
\ninept 
\maketitle

\begin{abstract}
In this paper, we present AISHELL-3, a large-scale and high-fidelity multi-speaker Mandarin speech corpus which could be used to train multi-speaker Text-to-Speech (TTS) systems. The corpus contains roughly 85 hours of emotion-neutral recordings spoken by 218 native Chinese mandarin speakers. Their auxiliary attributes such as gender, age group and native accents are explicitly marked and provided in the corpus. Accordingly, transcripts in Chinese character-level and pinyin-level are provided along with the recordings. We present a baseline system that uses AISHELL-3 for multi-speaker Madarin speech synthesis. The multi-speaker speech synthesis system is an extension on Tacotron-2 where a speaker verification model and a corresponding loss regarding voice similarity are incorporated as the feedback constraint. We aim to use the presented corpus to build a robust synthesis model that is able to achieve zero-shot voice cloning. The system trained on this dataset also generalizes well on speakers that are never seen in the training process. Objective evaluation results from our experiments show that the proposed multi-speaker synthesis system achieves high voice similarity concerning both speaker embedding similarity and equal error rate measurement. The dataset\footnote{\url{www.aishelltech.com/aishell_3}}, baseline system code and generated samples\footnote{\url{sos1sos2sixteen.github.io/aishell3}} are available online.
  
\end{abstract}

\begin{keywords}
open source database, Text-to-speech, multi-speaker speech synthesis, speaker embedding,  end-to-end 
	
\end{keywords}

\section{Introduction}
Speech synthesis, or Text-To-Speech(TTS), is the automated process of mapping input text specifications to target utterances \cite{taylor2009text}. In recent years, neural network based TTS synthesis systems have achieved marvelous results in terms of audio quality and perceptual naturalness \cite{shen2018natural}. This flourishing research progress is made largely due to the introduction of attention based sequence-to-sequence modeling architectures such as Tacotron \cite{shen2018natural,wang2017tacotron} or Transformer-TTS \cite{li2019neural}, and neural vocoders that map the lower dimensional acoustic representation to waveforms \cite{oord2016wavenet,prenger2019waveglow,kumar2019melgan}.

A key characteristic of TTS is the lack of constraint, which renders the task essentially a one-to-many mapping \cite{taylor2009text}. Since given only textual content, speeches uttered by either male or female, with voices agitated or neutral, are equally valid outputs. But real-world application of such systems requires robust and consistent behaviors. This begs the question of whether we could provide further specification to the system to gain more flexibility over conventional approaches. There is a growing interest within the field in designing TTS systems that are more flexible and admits stronger constraints on its behaviors. Recent publications on expressive or prosodic TTS systems tend to associate the acoustic model with explicit control signals (e.g., pitch/energy for supervised settings \cite{valle2020mellotron} and learned embeddings for unsupervised variants \cite{wang2018style, sun2020fully, sun2020generating}) as augmented input besides normalized texts. A more prominent and intuitive feature of speech is the speaker identity, and multi-speaker acoustic models give TTS systems the ability to disentangle perceptual speaker identity from the textual contents of the synthesized utterance by explicitly conditioning the model on the desired speaker \cite{gibiansky2017deep,jia2018transfer,cooper2020zero,cai2020speaker}. 

Training such systems naturally requires significant amount of annotated data. VCTK \cite{veaux2016superseded} is a freely available multi-speaker corpus that could be used to train such systems. However, VCTK only contains recordings in English. As suggested by previous studies \cite{pan2019mandarin, yan2020mandarin}, despite the cultural influence of English language as a \textit{lingua franca} in the academia, language specific subsystems and model modifications are indeed an area of active research. TTS systems targeted on tonal languages such as Chinese Mandarin and Japanese face difficult situations considering their complex tonal and prosodic structures \cite{minematsu2012improved}. The lack of a publicly available multi-speaker Mandarin dataset suitable for TTS system training makes researches in this area more difficult and costly, and lacks objective indicators that are comparable across studies.

To this end, we introduce the AISHELL-3 corpus in this paper to fill this vacancy in open resources. AISHELL-3 contains roughly 85 hours of high fidelity Mandarin speech recordings from 218 native speakers, with manually transcribed Chinese characters and pronunciations in the form of pinyin notation. Furthermore, we present a multi-speaker TTS system trained with this dataset as a baseline system. Objective evaluations on the synthesized samples show consistent behavior with previous studies conducted on a VCTK system with the same architecture.

\section{The AISHELL-3 Dataset}

The AISHELL-3 dataset is a multi-speaker Mandarin Chinese audio corpus, which could be used to train multi-speaker TTS systems. There are in total 88035 recordings from 218 native speakers reading off text from given scripts with neutral emotions. All utterances are recorded in a quiet indoor environment using high fidelity microphones (44.1 kHz, 16 bit depth), which are 20 cm away from the speaker. The topics of the textual content spread a wide range of domains including smart home voice commands, news reports and geographic information.

Each recording contains accompanying transcripts in both Chinese characters and pinyins, which is the official latinized notation for marking Chinese pronunciations. The pinyin transcripts are obtained through human listening tests and directly corresponds to the speaker's actual readings, which addresses four key difficulties in automatically deriving them from the Chinese characters via dictionary lookups:
1. \textbf{Homograph.} Some characters could be pronounced in multiple ways depending on the textual context they resides \cite{cai2019polyphone}.
2. \textbf{Tone sandhi.} Some tones shift under certain phonological contexts, a good example being, normally characters in the initial part of consecutive third tones shift to the second tone, e.g. \textit{guan3 li3} (to manage) should be pronounced \textit{guan2 li3}, but this rule does not apply to all such circumstances.
3. \textbf{Erization (erhua).} The Chinese character for \textit{son} (pronounced \textit{er2}) behaves like a normal character in some contexts, but it also acts as an erization marker indicating the preceding character has an \textit{erized} final. 
4. \textbf{Accents and Mispronunciations.}
The above difficulties make the phonetizing process non-trivial and the manually labeled transcripts valuable for obtaining a pure dataset for training high quality TTS models and further development of more sophisticated conversion systems.

\begin{figure}[h]
  \centering
  \includegraphics[width=0.45\textwidth]{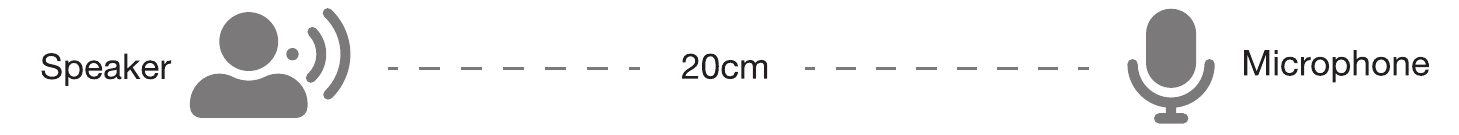}
  \caption{The recording environment setup}
  \label{fig:acoustic}
\end{figure}

\begin{table}[]
    \centering
    \caption{Utterance length \& speaker data-size distribution}
    \begin{tabular}{lcccc}
    \toprule
        &  mean & median & max & min  \\
    \midrule
    utterance length (character)   & 11.3    & 11   & 39    &1    \\
    \hline
    utterances by speaker   & 403.84   & 452    & 505   & 138    \\
    \bottomrule
    \end{tabular}
    \label{tab:len_dist}
\end{table}

\begin{table}
    \centering
    \caption{Speaker Attribute Distribution}
    \begin{tabular}{lc}
    \toprule
    attribute    & distribution \\
    \midrule
    gender  & 176 female / 42 male   \\
    \hline
    accent  & 165 north / 51 south / 2 others   \\
    \hline
    age group$^{*}$ & A:2 / B:173 / C:35 / D:8    \\
    \bottomrule
    \end{tabular}
    
    \begin{tablenotes}
    	\centering
	    \item[1]{* in years, A:$<16$, B:$16-25$,  C:$26-40$, D:$>41$}
    \end{tablenotes}

    \label{tab:attr_dist}
\end{table}

Table \ref{tab:len_dist} and table \ref{tab:attr_dist} show the distribution of some basic attributes across the entire dataset. It is worth mentioning that speaker genders are biased towards females, and the age distribution centers on young adults of around 20 years old.

\section{Baseline System}
The baseline system is a composition of three independently designed components: a textual-frontend, which performs necessary analysis and conversion on the textual input, an acoustic model, which maps textual and speaker specifications to a series of feature vectors, and a neural vocoder to decode the feature sequence into audio waveforms.

\subsection{The Speaker-agnostic Subsystem}
The textual-frontend and vocoder modules are implemented as speaker-agnostic considering the textual analysis and acoustic feature inversion processes involved in the TTS pipeline have lower correlations to speaker identities than that of the acoustic model. The textual-frontend, which is used to preprocess the datasets' raw labels, is composed of a pinyin-parser, a phonetizer, and a LSTM based prosody prediction model. We use a MelGAN \cite{kumar2019melgan} model trained with recordings from the AISHELL-3 dataset regardless of speaker labels as the neural vocoder.

In addition, we've found that Tacotron-2 alone, when naively trained with only Chinese phoneme or pinyin inputs, tends to produce synthetic samples with monotonic prosody. Considering the prosody have a high correlation with the naturalness in mandarin synthesis, we apply an RNN based prosodic label prediction model trained on a corpus labeled with prosodic taggings \cite{zhang2019prosodic}. Then the prosodic annotations are added into the phoneme sequence \cite{dictionary1998carnegie} as the final input sequence to the acoustic model.

The prosody label prediction task is formulated as a sequence tagging problem: the input sequence is a concatenation of learnable character embeddings, BMES-tagged \cite{ijcai2017-553} word segmentations and part-of-speech tags. The input sequence is first encoded with 2 layers of bi-directional LSTMs and then decoded by 2 fully-connected output-layers to produce BMES-tagged prosodic predictions. The prediction is made in a hierarchical fashion where the prosodic phrase is conditioned on wording predictions in light of the hierarchical nature of the structure of the prosodic units.

\subsection{The Speaker-aware Subsystem}
We follow a recently published work on multi-speaker TTS incorporating speaker-embedding feedback constraint \cite{cai2020speaker} to build our speaker-aware acoustic model. It is a Tacotron-2 based architecture with an extra speaker-encoder module as illustrated in figure \ref{fig:acoustic}.

\subsubsection{Tacotron-2 backbone}
Tacotron-2 follows the basic structure of a sequence-to-sequence architecture, which is composed of an encoder, a decoder and an attention mechanism in between. The encoder module of the tacotron-2 model is a CNN-BiLSTM based neural network that maps input texts into a sequence of hidden states. This hidden state sequence is then used as $values$ in an attention mechanism to calculate the context vector for every step of the decoder module. The attention mechanism used by Tacotron-2 is the hybrid attention, which is a form of stateful additive attention that leverages both content and location information of the decoding process. The decoder module of Tacotron-2 can be decomposed into three parts: prenet, postnet, and the decoder RNN. for every time-step in the decoding process, the mel-spectrum from the previous step is mapped to a lower dimensional hidden representation by prenet, which is a series of relu-activated dense layers. The processed frame and context vector are then passed to two layers of zone-out LSTMs \cite{krueger2016zoneout}, whose hidden-state is used by the attention mechanism to produce the current context vector. Finally, the hidden state and context vector is concatenated and fed into two projection layers to form the output acoustic feature frame for the current time-step and the \textit{stop-token}, which signals whether the last feature frame has been produced in inference stage. The postnet is itself two layers of 1d-convolution, and the final output feature frames are obtained by passing the decoded frames through postnet together with	 a residual connection.

\begin{figure}[h]
  \centering
  \includegraphics[width=0.46\textwidth]{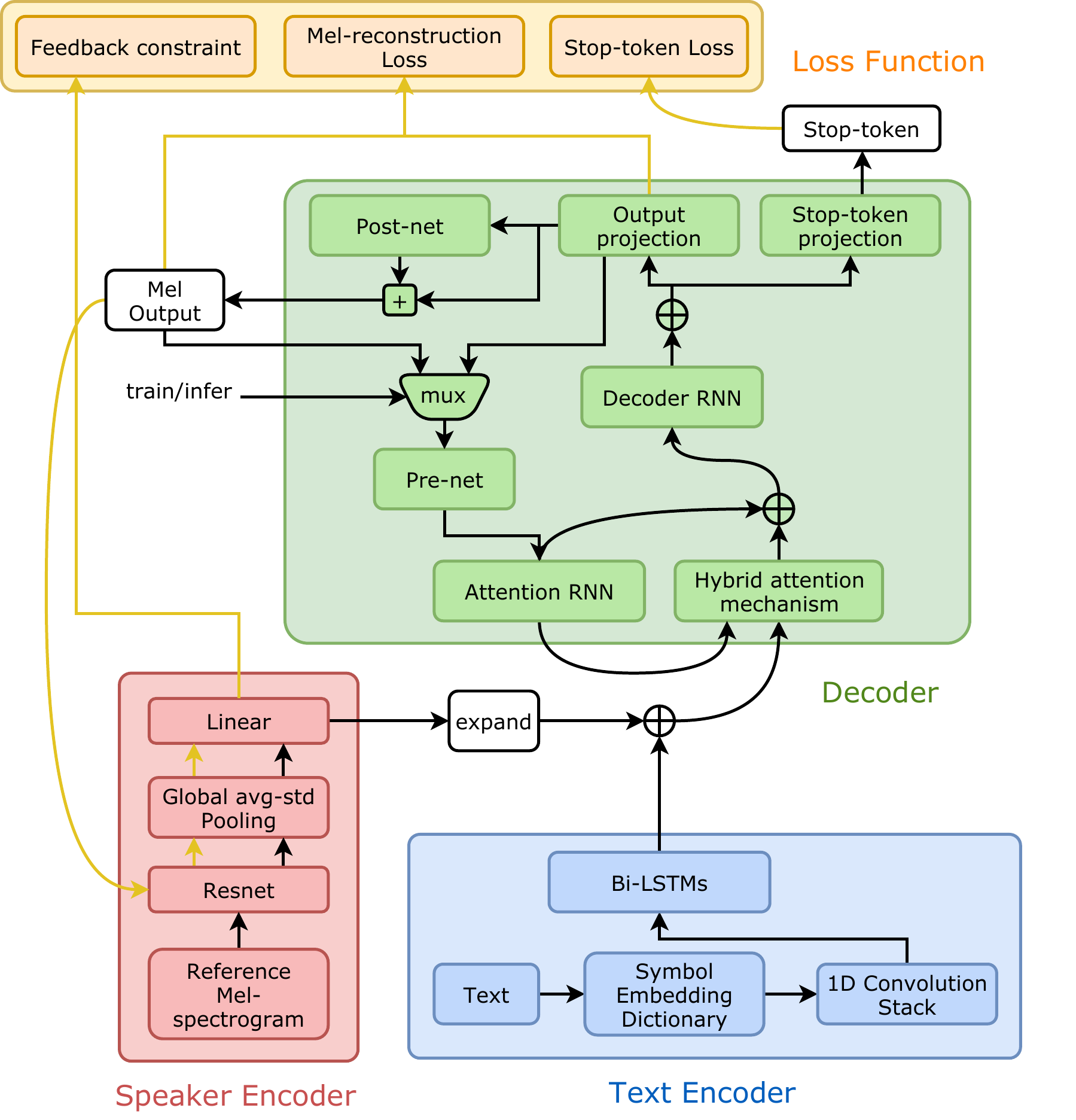}
  \caption{The speaker-aware subsystem}
  \label{fig:acoustic}
\end{figure}

\subsubsection{Speaker encoder}
The speaker encoder is used to extract discriminative speaker information from variable length reference audio inputs, which in our case is always the ground-truth audio during training. We follow \cite{cai2020fly} to implement a Resnet-based speaker verification network with global mean-std pooling as the speaker encoder for our baseline multi-speaker system. The speaker embedding vector used to condition our acoustic model is a linear-projection of the pooled mean and standard deviation feature vectors. 

\subsubsection{Feedback constraint}
We've learned from previous studies that concatenating speaker embedding vectors on encoder hidden states alone does not impose a strong enough incentive on the acoustic model to produce high fidelity results in terms of speaker similarity, especially when the reference audio is never seen in the training data \cite{cooper2020zero}. Recent work on speaker identity feedback constraint shows that by adding an additional loss term between reference embedding and the extracted embedding of the synthesized signal, the robustness and speaker similarity of the produced speeches are effectively increased \cite{cai2020speaker}. 
We follow the system description from the aforementioned study to modify our baseline multi-speaker acoustic model. As shown in Figure \ref{fig:acoustic}, each optimization step in the training process includes an additional speaker embedding cosine similarity loss term. The parameters in the speaker encoder module are pre-trained, and frozen during the training process of the acoustic model. The final loss function includes several terms:
$$ L = MSE_{mel} + BCE_{stop} - \alpha \cdot cos(E_{ref},E_{synth}) $$
Where $MSE_{mel}$ refers to the \textit{mean square error} between the predicted and ground-truth mel-spectrograms, the $BCE_{stop}$ term denotes the \textit{binary cross-entropy} on the stop-token, and the last term denotes the cosine similarity between synthesized and ground-truth speaker embeddings. The weight hyper-parameter $\alpha$ is set to $1.0$.

\section{Experiments}
We implemented and trained the baseline synthesis system and performed evaluations on the synthetic samples. The acoustic model and neural vocoder are trained using the presented dataset, and the speaker encoder module is trained separately using 775289 utterances from the aidataTang \cite{datatang} and MAGICDATA Mandarin \cite{magicdata} corpus. The detailed data preparation process for the acoustic model and evaluation results are described in this section.

\subsection{Data Preparation}

\subsubsection{Down-sampling}
To match the acoustic feature used for training the pre-trained speaker encoder, we downsample the data to 16kHz sample-rate, and uses window and hop lengths of 50 ms and 12.5 ms respectively.
\subsubsection{Train-test separation.} The training and testing data division is in broad terms made randomly based on speaker identities. Of the presented 218 speakers, 44 were drawn randomly from the population to form the unseen speakers test-set. The remaining 174 speakers are used to train the multi-speaker model described in chapter 3. It is worth noting that not all samples from the training set are used in model training, a portion of the utterances for each speaker in the training set are kept for seen speakers validation, The resulting train-set contains 64773 utterances, which is around 60 hours long.

\subsubsection{Silence trimming} 
In our proposed system, the attention alignment formed between encoder hidden states and decoder decoding steps directly controls the timing and rhythms of the synthesized utterance. However, since the alignment itself is an unsupervised by-product of the model optimization process, the stage in which a good alignment is formed could effectively affect the overall quality of the trained synthesis model. This correlation is demonstrated through experiments in \cite{battenberg2020location}. In our preliminary studies, we've found, by controlled experiments, that trimming the silence segments at the initial positions of train samples exceedingly speeds up the alignment formation during acoustic model training. So we performed energy-based VAD on all train samples' Mel-spectrogram and trimmed all initial silence segments. This led to an alignment formation speedup of 10 times in terms of optimization steps under 2 GTX-1080Ti GPUs.

\subsubsection{Long-form sentence augmentation}
It is observed that when the target sentence is too long during the inference phase, the synthesized utterance displays unstable attention alignment or unnatural speaking rate. This is due to the poor generalization ability of the hybrid-attention mechanism in long-formed synthesis commonly seen in Tacotron-2 models \cite{wang2018style,zheng2019forward}. Preliminary experiments show that by leveraging purely location-based attention mechanisms \cite{battenberg2020location}, this generalization problem is effected lifted at the cost of prosodic naturalness. However in our case, we turn towards data augmentation as a solution to this problem.

The average sentence length of the entire dataset is 11.3 characters, and a fair number of samples cover up to more than 20 characters long. We concatenate the acoustic feature and textual labels of $N$ randomly selected sentences drawn from the same speaker to produce a longer training sample, where $N$ is a discrete $r.v.$ drawn from the distribution $(P_2=0.6, P_3=0.2, P_4=0.2)$. Here $P_n$ denotes the \textit{probability mass function} $f_N(\cdot)$ of $N$. By this augmentation, we generated 30000 samples as the augmented training data. The augmented data is used to fine-tune the TTS model converged on the original dataset.

\subsection{Objective Evaluation}
We perform objective evaluations to assess synthetic samples produced by the baseline model in terms of speaker similarity. The evaluation is performed on two groups: the validation set that contains the same speakers as the training set, which reflects the model’s ability to mimic the voice of speakers seen in the training process, and the test-set, which shows how well the model generalizes on reference speakers that are unseen during training.

\begin{figure}[h]
  \centering
  \includegraphics[scale=0.20]{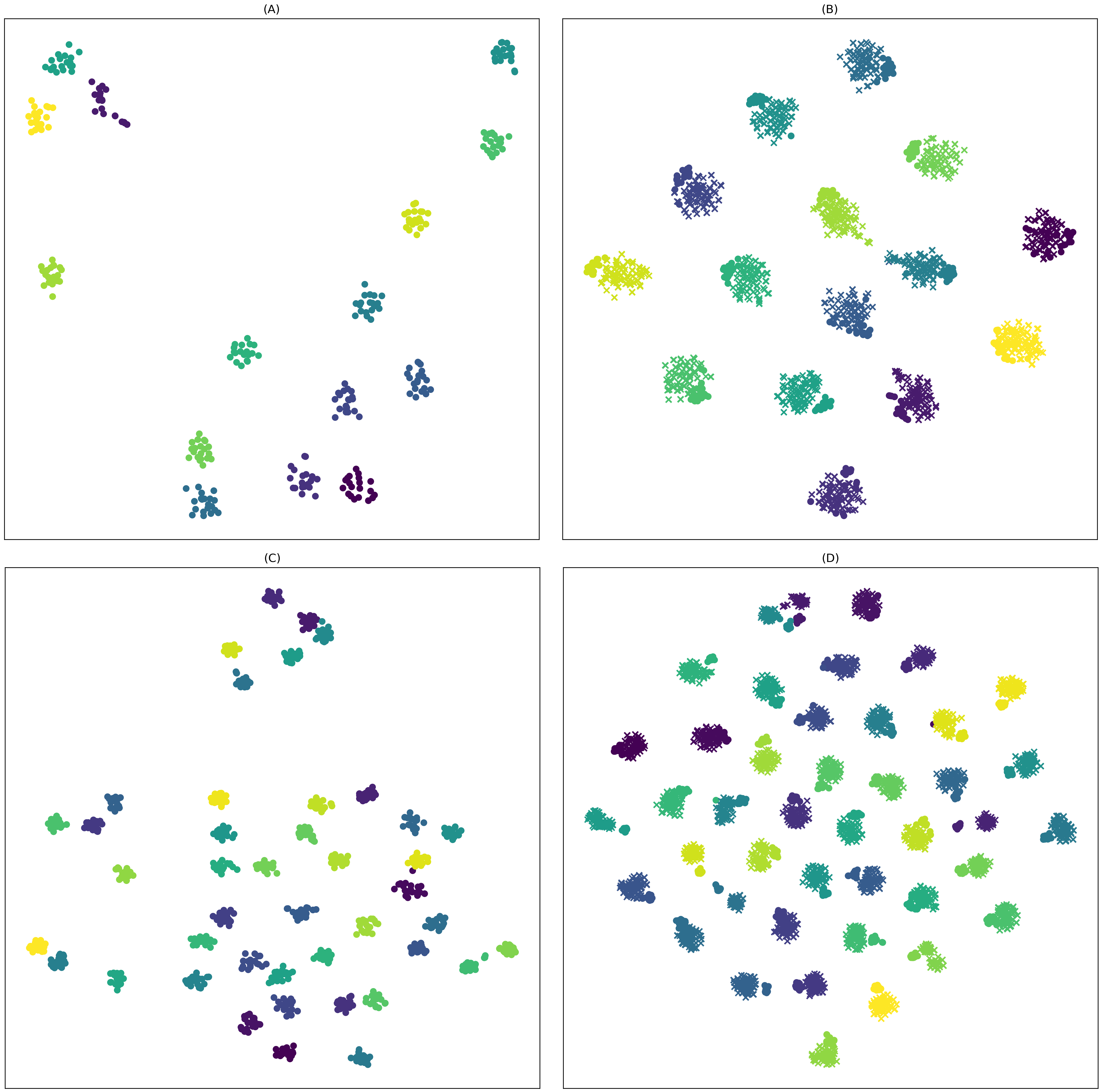}
  \caption{Speaker embedding visualization using t-SNE, embeddings extracted from audio recordings are represented with dots, synthesized with crosses. From left to right, top to bottom: (A) validation-set recordings; (B) validation-set recordings \& synthesized; (C) test-set recordings; (D) test-set recordings \& synthesized. Please note that the validation set contains seen speakers while the test-set only includes unseen ones.}
  \label{fig:vctk_textdep}
\end{figure}

In both trials, we synthesize 20 \textbf{text-dependent} and 20 \textbf{text-independent} utterances for every speaker, where text-dependent means the ground-truth audio with the same textual content is used to extract the speaker embedding, while text-independent is using the speakers' respective mean embedding vectors as the reference. The textual content of the synthesized samples are not seen in the training stage.
In addition, we synthesize 3 versions per each sample to account for the randomness introduced to the inference stage by the dropout layers in the prenet module. By assessing the resultant attention maps, we eliminate the mis-aligned samples. This is done by calculating cosine similarities of the attention score vectors between output steps, since a natural alignment would conform to a nearly diagonal trend, resulting in lower inter-step similarities.
 The speaker representation of each selected synthetic sample is extracted for the following measurements.

\textbf{T-SNE Plot.} We visualize the extracted speaker representation space as 2-D scatter plots using t-Distributed Stochastic Neighbor Embedding (t-SNE). Since t-SNE plot models relative distance between embedding vectors, clusters shown in A and C in figure \ref{fig:vctk_textdep} represent groups of utterances that are close in terms of speaker identities. We observe clearly that embeddings from the same speaker forms dense clusters that are well separated from other speaker-subspaces.

\begin{table}[]
    \centering
    \caption{Objective evaluation results for Baseline system}
    \begin{tabular}{lcc}
    \toprule
    Data pool    & \begin{tabular}[c]{@{}c@{}}SV-EER (\%) \\ Dep / Indep \end{tabular}   & \begin{tabular} [c]{@{}c@{}} Cosine distance \\ Dep / Indep \end{tabular} \\
    \midrule
    AISHELL-3 recording   & 4.47$^{*}$    & -    \\
    \hline
    AISHELL-3 validation   & 4.56 / 4.26    & 0.918 / 0.917    \\
    \hline
    AISHELL-3 test-set   & 9.46 / 9.56    & 0.868 / 0.871     \\
    \hline
    VCTK train-set   & 5.02 / 3.42    & 0.842 / 0.670   \\
    \hline
    VCKT test-set   & 8.22 / 7.68    & 0.764 / 0.577    \\

    \bottomrule
    \end{tabular}
    \begin{tablenotes}
    	\centering
	    \item[]{$^{*}$dep/indep not applicable}
    \end{tablenotes}

    \label{tab:my_label}
\end{table}

\textbf{Cosine Similarity.} We use cosine function to measure speaker embedding vector similarity, which is a common method used in speaker verification systems. We measured the cosine similarity between synthesized speechs' embeddings and embeddings extracted from ground truth audio samples. A higher value means higher similarity between synthesized and recorded samples as is considered by the speaker verification system.

\textbf{SV-EER.} We also make use of the concept of Speaker Verification (SV) Equal-Error-Rate (EER) as an objective evaluation index. EER is ubiquitously seen in SV tasks as a pointer to system performance, and it could also be interpreted as a measurement for the quality of a multi-speaker speech synthesis system. Perfect TTS systems would produce results that are indistinguishable from real data. To evaluate the system using the EER measure, we draw 10,000 pairs of samples from a pool of audio samples per trial. 

Table \ref{tab:my_label} presents our results and a similar experiment performed on the VCTK dataset published in \cite{cai2020speaker}. It’s observed that both validation-set and test-set speakers achieved high average cosine similarities between synthesized and real samples. And the EERs for validation speakers are only 0.65\% and 0.60\% higher than the SV-baseline. There is a performance drop of around 5\% for EER between test-set and validation-set speakers, which is consistent with results obtained on VCTK although the two speaker embedding networks are trained using different data.

\section{Conclusion}
\label{conclusion}
In this paper, a new publicly available Mandarin speech corpus that could be used to train multi-speaker TTS systems is presented. The dataset contains 88035 high-fidelity utterance recordings collected from 218 native speakers and includes hand-labeled full pinyin annotations. Moreover, a baseline multi-speaker TTS system based on speaker embedding feedback constraint is presented and described at length. We analyzed the characteristics of the presented dataset and incorporated a number of augmentation procedures, such as prosodic label prediction, silence trimming and long-form sentence concatenation to the data preparation process. The trained model is further evaluated in terms of speaker similarity and generalization capacity. 
And we found our presented multi-speaker TTS system is capable of producing natural speeches that imitates the voice of reference speakers. 
We draw the conclusion based on both objective and perceptual experimental results, that the presented corpus is valuable for the multi-speaker TTS research in Mandarin.

\noindent \textbf{Acknowledgements} This research is funded in part by the National Natural Science Foundation of China (61773413), Key Research and Development Program of Jiangsu Province (BE2019054), Six talent peaks project in Jiangsu Province (JY-074), Science and Technology Program of Guangzhou City (201903010040,202007030011).


\bibliographystyle{IEEEbib}

\bibliography{main.bib}

\end{document}